\documentclass[aps,twocolumn,amsmath,amssymb,showpacs,prl,superscriptaddress,unsortedaddress]{revtex4}
\usepackage{epsf}
\usepackage{graphicx}
\usepackage{gensymb}

\newcommand{\etal}{{\it et al.}}

\begin{document}

\title{Fermi surface and strong coupling superconductivity in single crystal NdFeAsO$_{1-x}$F$_x$}

\author{C. Liu}
\affiliation{Ames Laboratory and Department of Physics and
Astronomy, Iowa State University, Ames, Iowa 50011, USA}

\author{T. Kondo}
\affiliation{Ames Laboratory and Department of Physics and
Astronomy, Iowa State University, Ames, Iowa 50011, USA}

\author{M. E. Tillman}
\affiliation{Ames Laboratory and Department of Physics and
Astronomy, Iowa State University, Ames, Iowa 50011, USA}

\author{R. Gordon}
\affiliation{Ames Laboratory and Department of Physics and
Astronomy, Iowa State University, Ames, Iowa 50011, USA}

\author{G. D. Samolyuk}
\affiliation{Ames Laboratory and Department of Physics and
Astronomy, Iowa State University, Ames, Iowa 50011, USA}

\author{Y. Lee}
\affiliation{Ames Laboratory and Department of Physics and
Astronomy, Iowa State University, Ames, Iowa 50011, USA}

\author{C. Martin}
\affiliation{Ames Laboratory and Department of Physics and
Astronomy, Iowa State University, Ames, Iowa 50011, USA}

\author{J. L. McChesney}
\affiliation{Advanced Light Source, Berkeley National Laboratory,
Berkeley, California 94720, USA}

\author{S. Bud'ko}
\affiliation{Ames Laboratory and Department of Physics and
Astronomy, Iowa State University, Ames, Iowa 50011, USA}

\author{M. A. Tanatar}
\affiliation{Ames Laboratory and Department of Physics and
Astronomy, Iowa State University, Ames, Iowa 50011, USA}

\author{E. Rotenberg}
\affiliation{Advanced Light Source, Berkeley National Laboratory,
Berkeley, California 94720, USA}

\author{P. C. Canfield}
\affiliation{Ames Laboratory and Department of Physics and
Astronomy, Iowa State University, Ames, Iowa 50011, USA}

\author{R. Prozorov}
\affiliation{Ames Laboratory and Department of Physics and
Astronomy, Iowa State University, Ames, Iowa 50011, USA}

\author{B. N. Harmon}
\affiliation{Ames Laboratory and Department of Physics and
Astronomy, Iowa State University, Ames, Iowa 50011, USA}

\author{A. Kaminski}
\affiliation{Ames Laboratory and Department of Physics and
Astronomy, Iowa State University, Ames, Iowa 50011, USA}

\date{\today}
\begin{abstract}
We use angle-resolved photoemission spectroscopy (ARPES) to
investigate the electronic properties of the newly discovered
oxypnictide superconductor, NdFeAsO$_{1-x}$F$_x$. We find a
well-defined Fermi surface that
consists of a large hole pocket at the Brillouin zone center and a
smaller electron pocket in each corner of the Brillouin zone.
The overall location and shape of the Fermi surface agrees reasonably
well with calculations. The band dispersion is quite complicated with
many flat bands located just below the chemical potential. We
observe a superconducting gap of 20 meV, which indicates that this
system is in the strong coupling regime. The emergence of a
coherent peak below the critical temperature $T{_\textrm{c}}$ and
diminished spectral weight at the chemical potential above
$T{_\textrm{c}}$ closely resembles the spectral characteristics of
the cuprates.
\end{abstract}


\maketitle

Superconductivity is a spectacular manifestation of quantum
mechanics on a macroscopic scale. In most cases this phenomenon can be
well understood within the Bardeen-Cooper-Schrieffer theory as phonon
mediated pairing of electrons and condensation of the resulting boson
gas. All known superconductors that can be understood within this theory have critical temperatures,
(below which they become superconducting or $T{_\textrm{c}}$) less than $\sim$40K \cite{McMillanBCS}. For over twenty
years the only known exception was the cuprate family of high
temperature superconductors where superconductivity of unknown
mechanism exists up to $\sim$130K. Recently, the discovery
of superconductivity with a $T{_\textrm{c}}$ up to 55K in a new class
of materials with the general formulae LnFeAsO$_{1-x}$F$_x$ (Ln
being the lanthanides La, Sm, Ce, Nd, Pr, and Gd, \textit{x} being
the fluorine doping level) \cite{Kamihara_original, Takahashi43K,
Ren_55K} raises interesting questions on whether the mechanism of
pairing in these new materials is the same as in the cuprates, or if there
is yet another way superconductivity can be established.

There are already a number of different scenarios proposed for the
mechanism in this new class of materials. In some theories
the superconductivity arises due to the suppression of spin density
wave (SDW) ordering between the $\Gamma$ and M Fermi sheets
\cite{Dong_CompetingSDW, la_Cruz_NeutronScattering, Weng_SDW,
Ran_NodalSDW, Wang_Renormalization}. For this scenario to work both
sheets need to have similar radii. In other classes of theory, flat
bands just below the chemical potential are important as they can
couple electrons at the Fermi surface via Coulomb or Hund's rule
scattering.
A prerequisite to understanding the mechanism of superconductivity
in these materials is the knowledge of the low lying electronic
excitations because these are the states that eventually compose the
superconducting condensate. A number of angle integrated
photoemission studies are available in the literature
\cite{Sato_GapPE, Jia_PE, Liu_PE} on this subject, but as yet no
angle resolved data is available that addresses the questions about
the low lying electronic excitations.
Here we present data from angle resolved photoemission spectroscopy
(ARPES) on the Fermi surface, band dispersion and superconducting gap in
NdFeAsO$_{1-x}$F$_{x}$. We find that the Fermi surface consists of a
cylindrical hole pocket centered around $\Gamma$ (0,0,0) and a
cylindrical electron pocket at each corner of the Brillouin zone (M
points), similar to most calculations
\cite{Dong_CompetingSDW, Lebegue_FS_LaOFeP, Singh_FS, Xu_FS}. Some
band calculations yield also an ellipsoidal Fermi surface around Z
(0,0,1) \cite{Lebegue_FS_LaOFeP, Singh_FS}. We did not observe
this for the studied photon energies, however we did find a flat top
band just below the chemical potential centered at $\Gamma$. Even
with a very small $k_{z}$ dispersion, this band could cross the
chemical potential and give rise to the ellipsoidal Fermi surface at
Z.
We observe a superconducting gap of $\Delta$ = 20 meV at the
$\Gamma$ hole pocket for \textit{T} = 20K. This corresponds to
$2\Delta/T_\textrm{c}$ = 8 and indicates that this system is in the
strong coupling regime. We note that this ratio is similar to that
of the cuprate family of high temperature superconductors. The
emergence of the coherent peak below $T{_\textrm{c}}$ and diminished
spectral weight at the chemical potential above $T{_\textrm{c}}$
closely resembles the spectral characteristics of the cuprates.
\begin{figure*}
\includegraphics[width=6.5in]{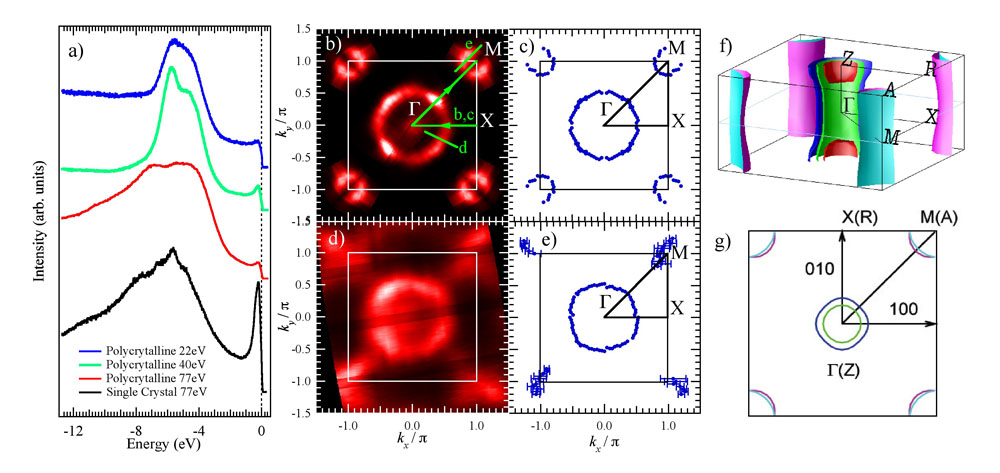}
\caption{(color) Measured and calculated Fermi surface of
NdFeAsO$_{1-x}$F$_x$ and NdFeAsO respectively. {\bf a}, Angle
intergrated spectra from polycrystalline and single crystal samples
for selected photon energies. {\bf b}, Fermi surface map - intensity
of the photoelectrons integrated over 20 meV about the chemical
potential obtained with 22 eV photons at \textit{T} = 70K. The areas
of bright color mark the locations of the Fermi surface. Green lines
and arrows mark the location of the selected cuts for Figure 2a-b.
{\bf c}, The locations of the Fermi crossings extracted from the raw
data by fits to the momentum distribution curves (MDCs). {\bf d},
same data as in b) but obtained with 77 eV photons. {\bf e}, same
data as in c) but obtained with 77 eV photons. {\bf f},
3-dimensional FS of NdFeAsO obtained from FLAPW calculations. {\bf
g}, A FS cross-section at $k_{z}=0$ (X-$\Gamma$-M plane) obtained by
FLAPW calculations.} \label{fig1}
\end{figure*}

High pressure synthesis of samples with the nominal composition of
NdFeAsO$_{0.9}$F$_{0.1}$ was carried out in a cubic, multianvil
press, with an edge length of 19 mm from Rockland Research
Corporation. Stoichiometric amounts of NdFe$_{3}$As$_{3}$,
Nd$_{2}$O$_{3}$, NdF$_{3}$ and Nd were pressed into a pellet with a
mass of approximately 0.5 g and placed inside of a BN crucible with
an inner diameter of 5.5 mm. The synthesis was carried out at a
pressure of 3.3 GPa.  The temperature was increased, over one hour,
from room temperature to 1350 - 1400$\celsius$ and then held there
for 8 hours before being quenched to room temperature.  The pressure
was then released and the sample was removed mechanically.  This
synthesis produced a high density pellet that contained large grains
(up to 300 $\times$ 200 $\mu$m in cross section) of superconducting
($T{_\textrm{c}}$ $\sim$ 53 K) NdFeAsO$_{0.9}$F$_{0.1}$ as well as
non-superconducting NdFeAsO. In addition there are inclusions of
FeAs and Nd$_{2}$O$_{3}$. Magneto optical measurements \cite{Ruslan}
indicate that on average the samples are over 50\% superconducting.
The single crystals were mechanically extracted from the pellet.
They were then mounted on an Al pin using UHV compatible epoxy and
cleaved \textit{in situ} yielding flat mirror-like surface. The data
was acquired at the PGM beamline of the Synchrotron Radiation
Center, using a Scienta 2002 analyzer and at the Advanced Light
Source using beamline 12.0.1 equipped with a Scienta 4000 analyzer.
The energy and angular resolution were set at 30 meV and 0.5 degree,
respectively. The beam profile on the sample was slightly
elliptical, with a mean diameter smaller than 100$\mu$m. The
photoelectron energy corresponding to the chemical potential was
determined by measuring spectra from polycrystalline aluminium in
electrical contact with the sample. Measurements carried out on
several samples yielded similar results for the band dispersion and
Fermi surface. Our full-potential linearized plane wave (FLAPW)
calculation \cite{FLAPW} used the local density approximation(LDA)
\cite{LDA}, and the experimental lattice constants \cite{Quebe} for
the undoped parent compound NdFeAsO. For the internal parameters,
since there is no available experimental data we take atomic
positions from PrFeAsO \cite{Quebe} and optimize As position by
total energy minimization that results in $z_{\textrm {As}}$
=0.639$c$.

Measurements on single crystals are essential for an accurate
determination of the electronic structure. The presence of multiple
phases in polycrystalline samples can often contaminate the pristine
spectra. As an example we plot in figure 1a a comparison of the
photoemission data from polycrystalline samples and angle integrated
measurements from single crystals. The strong intrinsic peak just
below the chemical potential is very strongly suppressed in the
polycrystalline samples. A common method for illustrating the
topology of the Fermi surface is to plot the ARPES intensity at the
chemical potential as a function of momentum \cite{HelenFS, JoelFS}.
In Fig. 1b,d we plot the ARPES intensity integrated within 20 meV
about the chemical potential from data obtained using 22eV and 77eV
photons. The brightest areas in these plots indicate high
photoelectron intensity and therefore approximate to the location of
the Fermi surface. To obtain more accurate information we have
extracted the exact locations of the Fermi momentum from momentum
distribution curves (MDCs) using a procedure described
elsewhere\cite{JoelFS} and plot them in panels 1c and 1e. Our data
show that indeed this system has a Fermi surface that consists of
two main contours: a larger one centered at $\Gamma$ (0,0) and a
smaller one centered at the corners of the Brillouin zone. For
comparison we plot the 3-dimensional (3D) Fermi surfaces based on
FLAPW band calculations for fluorine-free NdFeAsO in figure 1f. A 2D
cut at $\Gamma$ ($k_{z}=0$) is shown in figure 1g. We note that
there are both similarities and differences between the data and
calculations. Overall the calculation predicts Fermi surface sheets
centered at $\Gamma$ and M, in agreement with our measurements. The
main difference between theory and experiment is the number of FS
sheets and their relative size. According to calculation there
should be two sheets at each symmetry point. It is quite possible
that the separation between the sheets is smaller than the
calculated result and not sufficient to resolve in the experiment. A
second disagreement is in the size of the $\Gamma$-pocket and the
ratio of its radius to that of the M-pocket. This poses a
significant challenge for models relying on nesting between the
$\Gamma$ and M Fermi surface sheets and can be possibly reconciled
when strong electron correlations are taken into account.

\begin{figure}
\includegraphics[width=3.5in]{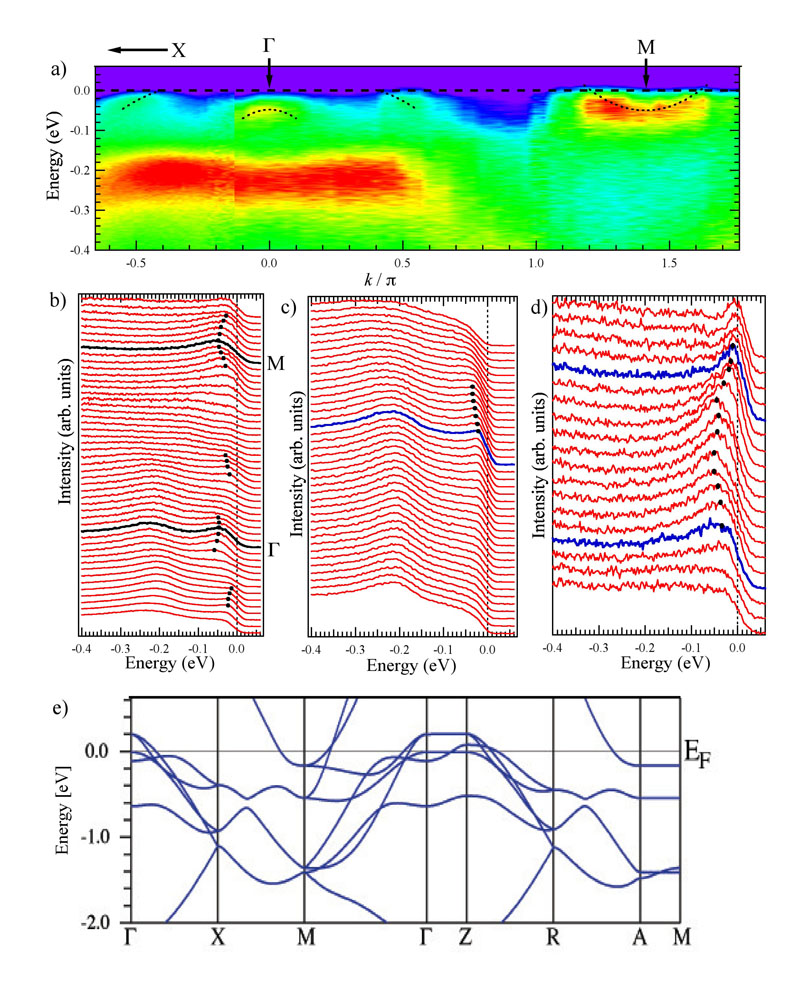}
\caption{(color) Measured and calculated band dispersion. Locations
of the cuts are indicated by green lines in figure 1(b). {\bf a},
Measured ARPES intensity along the X-$\Gamma$-M direction
(\textit{T} = 70K). Black dashed curves are guides to the eye . {\bf
b} Energy distribution curves (EDCs) for data in panel a) at
\textit{T} = 70K. {\bf c} and {\bf d}, EDCs (\textit{T} = 70K) in
the directions shown in figure 1b). Blue curves mark the Fermi
crossing momenta $k_{\textrm{F}}$. {\bf e}, Band dispersion along
the symmetry directions for fluorine-free NdFeAsO obtained from
FLAPW calculations.} \label{fig2}
\end{figure}

\begin{figure}
\includegraphics[width=3.5in]{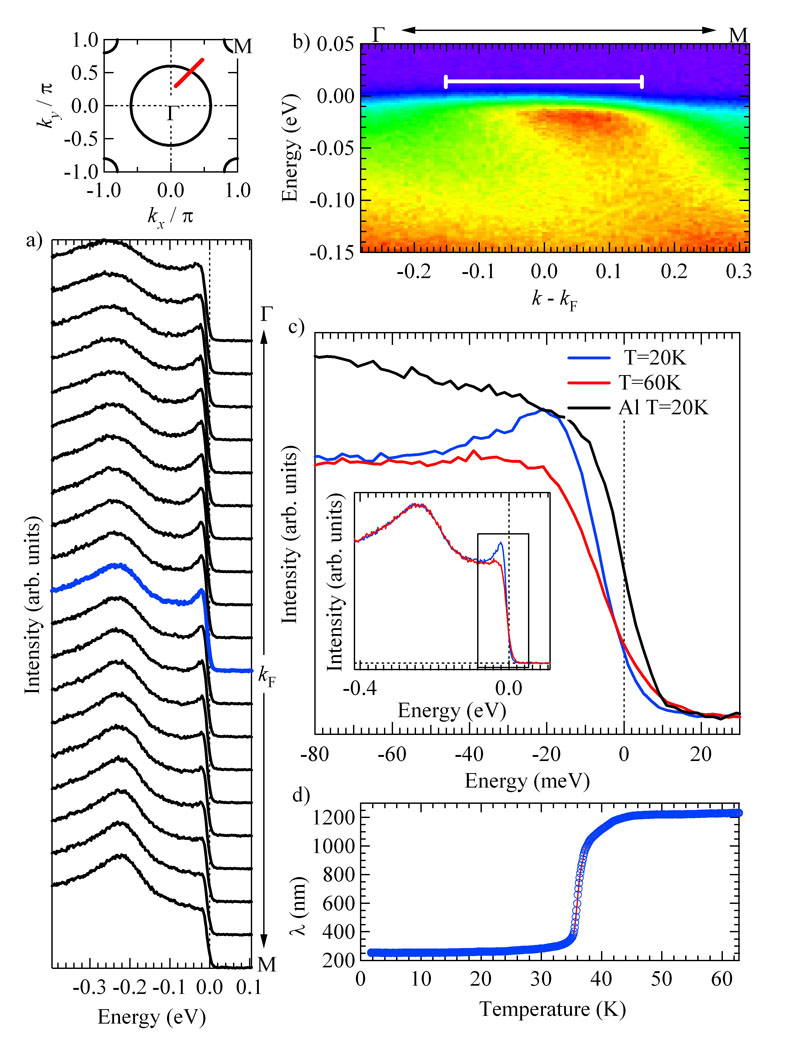}
\caption{(color) Superconducting gap at the $\Gamma$ pocket,
location marked in the upper-left inset. {\bf a}, EDCs along the
cut. Blue line indicates the EDC at the Fermi momentum. {\bf b},
Intensity map along this cut with a clearly visible dispersion and
back bending of the band close to Fermi energy $E_{\textrm{F}}$
caused by the superconductivity. The white line marks the range of
momenta for which the EDCs are plotted in panel a). {\bf c},
Magnified low binding energy region from inset, where EDCs at
\textit{T} = 60K and \textit{T} = 20K are plotted. Black curve is a
reference spectra from polycrystalline Al. {\bf d}, The penetration
depth data measured using the Tunneling Diode Resonator technique on
the same crystal after the ARPES measurement. A strong
superconducting signal is shown. } \label{fig3}
\end{figure}

The band dispersion along selected cuts (indicated by green arrows
in figure 1b) is shown in figure 2. Panel 2a shows the intensity map
along the X-$\Gamma$-M symmetry line. Bright areas on this graph
mark the locations of the bands. From these data we can identify the
topology of the main Fermi surface components. The Fermi surface
centered at $\Gamma$ is hole-like, i.e., the unoccupied states are
on the top of this band. In contrast, the Fermi surface centered at
M is electron-like with the bottom of the band located at $\sim$50
meV below the chemical potential. In addition to these two
conduction bands, there are also two very prominent fully occupied
bands for this particular value of $k_{z}$. The tops of both are
centered at $\Gamma$. The first is located just 50 meV below the
chemical potential. We speculate that this might be the band
responsible for the ellipsoidal pocket at Z, since even a small
amount of $k_{z}$ dispersion would bring it's top above the chemical
potential. The second band has a rather flat top which is located at
$\sim$200 meV below the chemical potential. This band is most likely
responsible for the strong peak observed in angle integrated
measurements \cite{Sato_GapPE, Jia_PE, Liu_PE}. EDCs for the above
data are shown in panel b. The dispersion of the peaks in this plot
agrees with the above description of the Fermi surface topology. The
band dispersion close to the chemical potential for the $\Gamma$
pocket and M pocket are plotted in panels 2c and 2d respectively.
The location of the Fermi momenta are marked by blue solid lines. In
panel 2e the calculated band structure at the high symmetry points
for fluorine-free NdFeAsO is presented for comparison.

We now proceed to discuss the measurements of the superconducing gap
in this material. The EDCs and ARPES intensity along a selected cut
at the $\Gamma$ pocket are shown in Fig. 3a-b. The Fermi momentum
was determined using the peak position of the MDCs and it is marked
in blue. The inset of panel (c) shows the EDCs below (red) and above
(blue) $T_{\textrm{c}}$, whereas in the enlarged panel we plot the
magnified part of EDCs close to the chemical potential. The black
curve is a chemical potential reference measured using
polycrystalline aluminium in electrical contact with sample. The
peak of the EDC above $T_{\textrm{c}}$ is very broad and its leading
edge is slightly shifted ($\sim$10 meV) towards higher binding
energies, which indicates the possible existence of a pseudogap. At
present we do not have extensive temperature dependent data to
definitely confirm whether these effects are similar to the
pseudogap observed in the cuprates. We have performed these
measurements with varying photon flux and confirmed that this shift
is not due to charging effects. Below $T_{\textrm{c}}$ the spectrum
changes quite dramatically - a coherent peak develops at a binding
energy of $\sim$20 meV which roughly corresponds to the value of the
superconducting gap. In addition, a characteristic back-bending of
the band is observed in the dispersion (panels 3a,b) which arises
due to particle-hole mixing in the superconducting state. The
spectral signatures of the superconducting transition are strikingly
similar to the ones widely reported in cuprates. Namely, below
$T_{\textrm{c}}$ a sharp peak appears, violating the conservation of
low energy spectral weight. This additional weight in the case of
the cuprates correlates reasonably well with the superfluid density
\cite{Fedorov, ShenCP}. Following the ARPES experiments we have
performed measurements of the temperature dependence of the
penetration depth on the same crystal. Due to the small size of the
crystals (200$\times$200$\times$30 $\mu$m), the measurements were
performed using a very sensitive Tunneling Diode Resonator technique
\cite{TDR}. Penetration depth data confirms that the crystals
measured by ARPES are indeed superconducting.

We are grateful for useful discussions with J\"{o}rg Schmalian. We
thank Helen Fretwell for useful remarks and corrections. Work at
Ames Laboratory was supported by the Department of Energy - Basic
Energy Sciences under Contract No. DE-AC02-07CH11358. The
Synchrotron Radiation Center is supported by NSF DMR 9212658. ALS is
operated by the US DOE under Contract No. DE-AC03-76SF00098.


\end{document}